\documentclass[12pt]{article}
\usepackage{amsmath,amssymb,amsfonts,amsthm}
\title{Gazeau$-$Klauder squeezed states associated with
    solvable quantum systems}

\author{M. K. Tavassoly
\\
\footnotesize{Atomic and Molecular Group, Faculty  of Physics, Yazd University, Yazd, Iran}
\\ \footnotesize{e-mail: mktavassoly@yazduni.ac.ir  } }

\begin{document}

\newcommand{\norm}[1]{\left\Vert#1\right\Vert}
\newcommand{\abs}[1]{\left\vert#1\right\vert}
\newcommand{\set}[1]{\left\{#1\right\}}
\newcommand{\R}{\mathbb R}
\newcommand{\I}{\mathbb{I}}
\newcommand{\C}{\mathbb C}
\newcommand{\eps}{\varepsilon}
\newcommand{\To}{\longrightarrow}
\newcommand{\BX}{\mathbf{B}(X)}
\newcommand{\HH}{\mathfrak{H}}
\newcommand{\D}{\mathcal{D}}
\newcommand{\N}{\mathcal{N}}
\newcommand{\la}{\lambda}
\newcommand{\af}{a^{ }_F}
\newcommand{\afd}{a^\dag_F}
\newcommand{\afy}{a^{ }_{F^{-1}}}
\newcommand{\afdy}{a^\dag_{F^{-1}}}
\newcommand{\fn}{\phi^{ }_n}
 \newcommand{\HD}{\hat{\mathcal{H}}}

 \maketitle

\begin{abstract}
 A formalism for the construction of some classes of Gazeau$-$Klauder squeezed
 states, corresponding to arbitrary solvable quantum systems with a known
 discrete spectrum, are introduced. As some physical applications, the proposed
 structure is applied to a few known quantum systems and then statistical
 properties as well as squeezing of the obtained squeezed states are studied.
 Finally, numerical results are presented.
 \end{abstract}


\section{Introduction}\label{sec-intro}
    The standard coherent states may be obtained from the action of
    the "displacement operator"(or "coherence operator") on the vacuum,
\begin{equation}\label{displace}
   D(z)=\exp(  z a^\dag - z^* a),
   \qquad D(z) | 0\rangle=|z  \rangle,
\end{equation}
   where $a$, $a^\dag$ are the standard bosonic annihilation, creation
   operators, respectively.
   Nowadays generalization of coherent states  and their experimental generations,
   have made  quantum physics much interesting, especially quantum optics \cite{AAG-book}.
   These states exhibit some interesting  {\it "nonclassical
   properties"} particularly quadrature squeezing, antibunching,
   sub-Poissonian photon statistics and oscillatory number distribution.

   Although some classes of generalized coherent states may possess
   squeezing in one of the quadrature components of the radiation field,
   another set of states known as
   {\it "squeezed states"}  which are the simplest
   representatives of nonclassical states also play an important role in quantum optics.
   These states are nonclassical states
    of the electromagnetic radiation field in
   which certain observables exhibit fluctuations less than the
   vacuum. Among them the "standard squeezed states"
   obtained by the action of "squeezing operator" on the vacuum \cite{Stoler},
\begin{equation}\label{ss}
   S(\xi)=\exp\left[\frac 1 2 (\xi {a^\dag}^2 -\xi^* a^2)
   \right], \qquad S(\xi) | 0\rangle=| \xi\rangle.
\end{equation}
   Bearing in mind that a special realization of the $su(1, 1)$ algebra
   can be considered with the
   generators $K_0=\frac 1 2 (a^\dag a + \frac 1 2 ),\; K_+ = \frac 1 2 (a^\dag) ^2,\;
   K_- = \frac 1 2 a^2$, the squeezed operator $S(\xi)$ in (\ref{ss}), may be
   re-written as the displacement operator
   $D(\xi)=\exp \left[\frac 1 2 ( \xi K_+ - \xi^* K_- )\right]$.
   These are the Perelomov \cite{Perelomov1} form of generalized coherent states
   which defined for reducible representation of
   $SU(1, 1)$ group in analogy with the displacement operator definition.
   Therefore, the squeezed states in (\ref{ss}) sometimes have been called
   the $SU(1, 1)$ coherent states,
   which are of great relevance to quantum optics for single mode fields.
   The number states expansion of the states in (\ref{ss})
   will be re-obtained as a special case
   of the proposed formalism in the present manuscript  (see example 1 in section 4).
   Anyway, squeezing (of the quantized radiation field)
   means that  the  uncertainty of one of the
   quadratures of the field falls below the uncertainty of the vacuum state
   at the cost of increased uncertainty in the other quadrature, hence
   the state is nonclassical.
   The usefulness of this property in various fields such as
   the measurement techniques
   and detection of gravitational waves \cite{Caves},
   enhancement and suppression of spontaneous emission
   \cite{Carmichael},
   and optical communication \cite{Shapiro}  are well understood.
   Some generalizations of the squeezed states are also introduced in
   literature. To say a few, it may be referred to as {\it "squeezed coherent
   states"}, {\it "nonlinear squeezed states"} \cite{Kwek},
   representations of squeezed states in an $f$-deformed Fock space \cite{rokntvs}
   and {\it "a class of nonlinear squeezed states"}
   recently introduced in \cite{Obada} (for a squeezed review on nonclassical states
   see \cite{Dodonov} and references therein).

   The present work is motivated to enlarge the class of
   squeezed states and especially provide a framework to be able to introduce
   the "squeezed states" in a direct relation to  physical systems.
   Recall that this purpose has been achieved in the
   "generalized coherent states" domain,
   by J-P. Gazeau and J. R. Klauder in an elegant
   fashion \cite{gazklau, Klauder96, Klauder98}.
   Actually, in addition to the {\it "continuity"} and the {\it "resolution of the identity"},
   they imposed {\it "temporal stability"} and {\it "action identity"} requirements
   as two new physical criteria, to force the generalized coherent states to
   more classical situation.
    Obviously,
    any (coherent or squeezed)  state preserves the temporal
    stability under the Hamiltonian dynamics, if one considers the eigen-value equation
    $\hat H |n\rangle = n |n\rangle$, i.e. that of harmonic oscillator.
    Whereas, if one  deals with a specific quantum
    system with Hamiltonian $\HD$ and eigen-energies $e_n$,
    so that $\HD |n\rangle = e_n |n\rangle$,
    and looks forward to construct the coherent or squeezed states,
    there will appear some problems with temporal stability of the states in hand.
    Fortunately,
    the proposed formalism of Gazeau and Klauder attacked the problem
    and the so-called {\it "Gazeau$-$Klauder coherent states"} possess the temporal stability
    property, under the action of the time evolution operator $e^{-i \HD t}$, essentially.
    The Hamiltonian $\HD$ in the latter operator
    is responsible for the dynamics of the quantum
    system. As  Roknizadeh {\it et al} established algebraically
    in \cite{Roknizadeh2004, Roknizadeh-Tav-AIP},
    the Hamiltonian $\HD$ is constructed by $A_{\rm GK}=af_{\rm GK}(\alpha, \hat n)$ and
    $A^\dag=f_{\rm GK}^\dag(\alpha, \hat n)a^\dag$
    as  the $f$-deformed annihilation and creation operators, respectively,
    via the factorization method
    $\HD = A_{\rm GK}^\dag A_{\rm GK}$. In this fashion, the Gazeau$-$Klauder coherent states
    are the generalized nonlinear coherent states with the
    generalized operator valued nonlinearity
    function $f_{\rm GK}(\alpha, \hat n)$,
    which depends explicitly on the intensity of light.

    Finally, a point is worth to mention.
    Recall that in the construction of the dual pair of
    Gazeau$-$Klauder coherent states in \cite{Roknizadeh-Tav-AIP}
    a different and non-usual method has been employed.
    This was due to recognition  that the
    dual pair of the Gazeau$-$Klauder coherent states must possess all of the
    four mentioned criteria in  \cite{gazklau}, carefully.
    But, as it will be demonstrated in the continuation of the present manuscript,
    although the presented formalism is essentially based on
    the structure of Gazea$-$Klauder coherent states,
    it is not necessary to reconsider  the whole criteria of
    Gazaeu and Klauder for the squeezed states will be introduced in this manuscript.
    The reason is clear, since when one deals
    with the  squeezed states, he(she) automatically relaxes from classicality.


   This paper is organized as follows. Section 2 is devoted to
   a brief review on the fundamental structure of Gazeau$-$Klauder coherent states. A new (and more general) proposal
   Gazeau$-$Klauder squeezed states associated with solvable quantum systems 11585
   for constructing a set of squeezed states, which have been called Gazeau$-$Klauder squeezed
   states', will be presented in section 3. Then, in section 4 the formalism will be applied to
   some quantum systems with a known discrete spectrum, and finally in section 5 the quantum
   statistical properties and squeezing of the obtained states will be studied.

\section{Gazeau-Klauder Coherent States as Nonlinear Coherent States}\label{sec-nl}

   Gazeau$-$Klauder  coherent states have been introduced associated with quantum systems with
   known discrete spectrum $E_n$ \cite{gazklau}.
   According to \cite {ElKinani2001b, gazklau}, the analytical
   representations of these states have been introduced as
   follows:
\begin{equation}\label{GKED}
    |z, \alpha \rangle \doteq \N(|z|^2)^{-1/2}
    \sum_{n=0}^{\infty}\frac{z^{n}e^{-i \alpha e_n}}{\sqrt{
    \rho(n)}}|n\rangle,  \qquad  z \in \C, \qquad 0 \neq \alpha \in \R,
\end{equation}
    where $\N(|z|^2)$ is some appropriate normalization constant.
    In equation (\ref{GKED}) the kets $\left\{|n\rangle \right\}_{n=0}^\infty$
    are the eigen-vectors of the Hamiltonian $\HD$,  with the
    eigen-energies $E_n$ such that,
 \begin{equation}\label{kps-14}
   \HD |n\rangle=E_n|n \rangle \equiv \hbar \omega e_n |n
   \rangle \equiv e_n|n\rangle, \quad \hbar = 1 =  \omega,
    \quad  n=0,1, 2, ... \;,
 \end{equation}
   where the re-scaled spectrum $e_n$, satisfy the inequalities
    $0=e_0 < e_1 < e_2 < \cdots < e_n < e_{n+1} < \cdots
    \;.$
    The action identity criteria with the condition $e_0=0$, imposed the requirement
    $$\rho(n)=[e_n]!, \qquad \Leftrightarrow \qquad e_n= \frac {\rho(n)}{\rho(n-1)}.$$
    It is established in \cite{Roknizadeh2004} that the states in the expansion (\ref{GKED})
    are {\it "nonlinear coherent states"} with the operator valued
    (and also intensity dependent) nonlinearity function
  \begin{equation}\label{nlGKED}
    f_{\rm {GK}}(\alpha, \hat{n}) =
    e^{i\alpha(\hat{e}_n-\hat{e}_{n-1})} \sqrt {\frac {\hat e_n}{\hat
    n}}, \qquad \hat e_n= \frac{\rho(\hat n)}{\rho(\hat n-1)},
  \end{equation}
    where $\hat n = a^\dag a$ is the number operator.
    The explicit dependence of the nonlinearity function on the
    spectrum of an arbitrary quantum system is notable.
    Therefore,  the  rising and lowering operators related to any
    solvable system may be defined as:
 \begin{equation}\label{crea-anni}
    A_{\rm GK}=a f_{\rm GK}(\alpha, \hat{n}),  \qquad A^\dag_{\rm GK}=
    f^\dag_{\rm GK}(\alpha, \hat{n})a^\dag,
 \end{equation}
    with a commutator between $A_{\rm GK}$ and $A_{\rm GK}^\dag$ as
 \begin{equation}\label{}
    [A_{\rm GK},A_{\rm GK}^\dag]=
    (\hat{n}+1)f_{\rm GK}(\hat{n}+1)f_{\rm GK}^\dag(\hat{n}+1)-
    \hat{n}f_{\rm GK}^\dag(\hat{n})f_{\rm GK}(\hat{n}).
\end{equation}
    Equations (\ref{nlGKED}) and (\ref{crea-anni}) show clearly that
    the $f-$deformed ladder operators for any solvable quantum system
    may be easily obtained.
    Using the {\it "normal-ordered"} form of the Hamiltonian
    and taking $\hbar=1=\omega$,
    for the Hamiltonian corresponding to Gazeau$-$Klauder coherent states  one
    gets
  \begin{equation}\label{normalH}
    \HD \equiv A_{\rm GK}^\dag A_{\rm GK} = \hat{n}
    \Big|f_{\rm GK}(\alpha, \hat{n})\Big|^2 \equiv
    \hat{e_n}.
  \end{equation}
   Consequently, the relation (\ref{kps-14}) holds, obviously.
   Moreover, one can get the two canonical
   conjugate of the operators $A_{\rm GK}$ and $A_{\rm GK}^\dag$,
   as
    $$B_{\rm GK}^\dag=\frac{1}{f_{\rm GK}^\dag(-\alpha, \hat
    n)}a^\dag, \qquad
    B_{\rm GK}=a\frac{1}{f_{\rm GK}(-\alpha, \hat n)},$$
     respectively.
   As a result
   $[A_{GK}, B^\dag_{GK}]=\hat I =[B_{GK},A^\dag _{GK}],$
    where $\hat I$ is the unit operator.
   Roknizadeh {\it et al} have introduced these set of operators to
   establish that the Gazeau$-$Klauder coherent states may be constructed
   by a non-unitary displacement type operator \cite{Roknizadeh-Tav-AIP}:
  \begin{equation}\label{NLCSs}
    D(z)|0\rangle = \exp(z A_{\rm GK}^\dag-z^*B_{\rm GK})|0\rangle=|z,
    \alpha\rangle.
  \end{equation}
    It is also shown that another displacement operator may be
    introduced, by which one can derive another
    class of nonlinear coherent states (have been called
    the {\it"dual states"} \cite{AliRokTav,  Roknizadeh-Tav-AIP}) as follows,
  \begin{equation}\label{DNLCSs}
    \tilde D(z)|0\rangle =
     \exp(z B_{\rm GK}^\dag-z^*A_{\rm GK})|0\rangle=|\widetilde{z,
      \alpha}\rangle.
  \end{equation}
    For achieving this purpose, recall that  the "dual family of Gazeau$-$Klauder
    coherent states"
    have been introduced in \cite{Roknizadeh-Tav-AIP} as follows,
  \begin{equation}\label{DGKCS}
      |\widetilde{z, \alpha} \rangle \doteq \N(|z|^2)^{-1/2}
      \sum_{n=0}^{\infty}\frac{z^{n}e^{-i \alpha \varepsilon_n}}{\sqrt{
       \mu(n)}}|n\rangle, \qquad \mu(n)=\frac{(n!)^2}{\rho(n)}, \quad z \in \C, \qquad  0 \neq \alpha \in \R.
  \end{equation}
     The deformed annihilation and creation operators $\widetilde{A}_{\rm GK}$
     and $\widetilde{A}^\dag_{\rm GK}$ of the dual oscillator algebra,
     encountered the operator valued nonlinearity function:
  \begin{equation}\label{nlDGKED}
     \widetilde{f}_{\rm GK}(\alpha, \hat{n}) = e^{i\alpha(\hat{\varepsilon}_n -
     \hat{\varepsilon}_{n-1})}\sqrt{ \frac {\hat \varepsilon_n} {\hat n}},
     \qquad {\varepsilon}_n \equiv \frac {\mu({n})}{\mu({n}-1)} \; .
  \end{equation}
     Hence, the deformed annihilation and creation operators
     of the dual oscillator may be expressed as:
  \begin{equation}\label{ladder-DGKED}
     \widetilde{A}_{\rm GK}= a \widetilde{f}_{\rm GK}(\alpha,
     \hat{n}),\qquad
     \widetilde{A}_{\rm GK}^\dag=  {\widetilde{f}_{\rm GK}}^\dag(\alpha,
     \hat{n})a ^\dag.
  \end{equation}
    The normal-ordered Hamiltonian of dual oscillator is therefore:
  \begin{equation}\label{Hamilt1}
    \widetilde{\HD} =
    \widetilde{A}^\dag_{\rm GK}\widetilde{A} _{\rm GK} =
    \hat n \left| \widetilde{f}_{\rm GK}(\alpha, \hat{n}) \right|^2
    \equiv  \hat \varepsilon_n.
 \end{equation}
    As a result,
 \begin{equation}\label{Hamilt2}
   \widetilde{\HD} |n\rangle =
    \varepsilon _n |n\rangle,
     \qquad  \varepsilon_n \equiv \widetilde{e}_n=\frac{n^2}{e_n},
 \end{equation}
    where again the units $\omega=1=\hbar$ have been used.
    The eigenvalues of $ \widetilde{\HD}$ are also required to satisfy
    the following inequalities:
    $ 0 = \varepsilon_0 < \varepsilon_1 <  \varepsilon_2< \cdots <
    \varepsilon_n < \varepsilon_{n+1} <   \cdots .$

   Now constructing the two conjugate operators of
   $\widetilde{A}_{GK}$ and $\widetilde{A}^\dag_{GK}$ i.e.
   $\widetilde{B}_{GK}=a\frac{1}{\widetilde{f}_{\rm GK}(-\alpha, \hat
   n)}$, $
   \qquad \widetilde{B}^\dag_{GK} =
   \frac{1}{\widetilde{f}^\dag_{\rm GK}(-\alpha, \hat n)}\;a^\dag, $
   respectively, one has
$   [\widetilde{B}_{GK}, \widetilde{A}^\dag_{GK}] = \hat I=
   [\widetilde{A}_{GK}, \widetilde{B}^\dag_{GK}].$

    Substituting $\alpha=0$, so that $e_n =  n f^2(n)$
    ($\varepsilon_n  = \frac{n}{f^2(n)}$) in the
    above relations (equations (\ref{GKED})  to (\ref{Hamilt2}))
    in this section will recover the  nonlinear
    coherent states (their dual family) which have  been already introduced
    by Man'ko {\it et al} \cite{Manko1997}, Matos Filho {\it et al} \cite{Matos1996, MMatos1996}
    (Ali {\it et al} \cite{AliRokTav} and Roy {\it et al} \cite{royroy}).

\section{ General Structure of Gazeau$-$Klauder  Squeezed States}\label{sec-nl}

   Following the explained path for the introduction of the
   dual pair of the Gazeau$-$Klauder coherent states,
   {\it "dual families of Gazeau$-$Klauder squeezed states"}
   for arbitrary solvable quantum systems with known
   discrete spectrum can be simply obtained.
   Upon using the considerations outlined in section 2,
   two new classes of  squeezed states may be introduced
   by the actions of the {\it "generalized squeezing operators"}
   $S$ and $\widetilde{S}$, which are now {\it "energy dependent"},
   on the vacuum as follows:
\begin{equation}\label{sop1}
   S(\xi, \alpha, f)|0\rangle =\exp\left[\frac 1 2(\xi {A_{\rm GK}^\dag}^2-\xi^*
   B_{\rm GK}^2)\right]|0\rangle = |\xi, \alpha, f\rangle,
\end{equation}
and
\begin{equation}\label{sop2}
   \tilde S(\xi, \alpha, f)|0\rangle = \exp\left[\frac 1 2 (\xi
   {B_{\rm GK}^\dag}^2-\xi^* A_{\rm GK}^2)\right]|0\rangle =
   |\widetilde{\xi,  \alpha, f}\rangle.
\end{equation}
   The explicit form of the generalized squeezed states, $|\xi, \alpha,
   f\rangle$, may be straightforwardly found by the superposition of even Fock states,
\begin{equation}\label{SS1}
   |\xi, \alpha, f\rangle = \N
    \sum _{n=0}^\infty \frac{\sqrt{(2n)!}}{n!} {[f\;^\dag_{\rm GK}(\alpha,
     2n)]!}
      \left[\frac{\exp(i \phi)\tanh r}{2}\right]^n|2n\rangle,
\end{equation}
   where in this case and what follows, $\xi= \tanh r \exp(i \phi)$
   and $\N$ is choosed so that the states be normalized
   (in what follows we will calculate it, explicitly).
   Inserting the explicit form of the nonlinearity function $f_{\rm GK}$ from
   (\ref{nlGKED}) with the help of the definition of Jackson's
   factorial one immediately  gets
$[f^\dag_{\rm GK}( \pm \alpha,
   n)]!= e^{ \mp i \alpha
   e_{n}}\sqrt{ \frac{[e_n]!}{n!}}\;.$
    To this end, the explicit form
   of the first class of the {\it "Gazeau$-$Klauder squeezed
   states"} will be expressed as the following superposition:
 \begin{equation}\label{SS11}
  |\xi, \alpha, f\rangle = \N \sum _{n=0}^\infty
   e^{-i \alpha e_{2n}}    \frac {\sqrt{[e_{2n}]!}  }{n!}
    \left[\frac{\exp(i \phi)\tanh r}{2}\right]^n|2n\rangle,
 \end{equation}
  where
 \begin{equation}\label{norm}
   \N= \left[ \sum_{n=0}^\infty \frac {[e_{2n}]!}{(n!)^2}
    \left(\frac{\tanh r}{2}\right)^{2n} \right]^{- \frac{1}{2}}.
 \end{equation}
  The states $|\widetilde{\xi, \alpha, f\rangle}$  introduced previously in (\ref{sop2}),
  the dual pair of the  generalized squeezed states in (\ref{sop1}),
  may be given by the superposition of even Fock states,
 \begin{equation}\label{DSS-Kwek}
   |\widetilde{\xi, \alpha, f}\rangle = \N'
    \sum _{n=0}^\infty \frac  {\sqrt{(2n)!}} {n!}\frac{1}{[f^\dag_{\rm GK} (-\alpha,
     2n)]!}
      \left[\frac{\exp( i \phi)\tanh r}{2}\right]^n |2n\rangle.
 \end{equation}
   Again inserting the equivalent form of the $[f^\dag_{\rm GK}(- \alpha,
   n)]!$ from (\ref{fact}) in terms of the eigenvalues of the system $e_n$,
   the { \it "dual family of Gazeau$-$Klauder squeezed states"}
   can be rewritten in the form
 \begin{equation}\label{SS22}
   |\widetilde{\xi, \alpha, f}\rangle =
    \N' \sum _{n=0}^\infty e^{- i \alpha e_{2n}} \frac{(2n)!}{n!\;\sqrt{[e_{2n}]!}}
     \left[\frac{\exp(i \phi)\tanh r}{2}\right]^n|2n\rangle,
 \end{equation}
    where
 \begin{equation}\label{normp}
    \N'=\left[ \sum_{n=0}^\infty \left( \frac {(2n)!}{n!} \right)^2 \frac {1} {[e_{2n}]!}
     \left( \frac {\tanh r}{2}\right)^{2n}
      \right]^{-\frac 1 2}.
 \end{equation}

   In addition to the above two distinct classes of  squeezed states
   (introduced in (\ref{SS11}) and (\ref{SS22})) which are in duality,
   it is also possible to propose two new sets of  squeezed states based on
   the dual Hamiltonian in (\ref{Hamilt1}) as follows,
  \begin{equation}\label{DSS1}
    S(\xi, \alpha, \widetilde f )|0\rangle
    = \exp \left[ \frac 1 2 \left(\xi ({{{\widetilde A}_{\rm GK} }^{\dag} })^2-\xi^\ast
     {\widetilde B}_{\rm GK}^2 \right) \right] |0\rangle =
     |\xi, \alpha, \widetilde f\rangle,
  \end{equation}
     and
  \begin{equation}\label{DSS2}
    \tilde S(\xi, \alpha, \widetilde f )|0\rangle = \exp\left[\frac 1 2 \left(\xi
    ({{\widetilde B}_{\rm GK}^\dag})^2-\xi^* {\widetilde
    A}_{\rm GK}^2\right)\right]|0\rangle=|\widetilde{\xi, \alpha, \widetilde f}\rangle.
  \end{equation}
   The generalized squeezed states, in (\ref{DSS1}) and (\ref{DSS2}), can be
   decomposed by the even Fock states as
 \begin{equation}\label{DSS11}
   |\xi, \alpha, \widetilde f\rangle = \widetilde \N
   \sum _{n=0}^\infty \frac{\sqrt{(2n)!}}{n!} [\widetilde f\;^\dag_{\rm GK}(\alpha,
   2n)]! \left[\frac{\exp(i \phi)\tanh r}{2}\right]^n|2n\rangle,
 \end{equation}
   and
 \begin{equation}\label{DSS3}
   |\widetilde{\xi, \alpha, \widetilde f}\rangle = {\widetilde\N'}
   \sum _{n=0}^\infty \frac{\sqrt{(2n)!}}{n!}
   \frac{1}{[\widetilde f\;^\dag_{\rm GK} (-\alpha, 2n)]!}
   \left[\frac{\exp(i \phi)\tanh r}{2}\right]^n |2n\rangle.
 \end{equation}
  Substituting the explicit form of
  $[\widetilde f\;^\dag_{ \rm GK}(\pm\alpha, n)]!=
  e^{\pm i \alpha \varepsilon_n}\sqrt{\frac{[\varepsilon_n]!}{n!}}$
  in (\ref{DSS22}) and (\ref{DSS3}),  one readily
  obtains the third and forth classes of
   {\it Gazeau$-$Klauder squeezed states} with the following superpositions:
 \begin{equation}\label{DSS22}
   |\xi, \alpha, \widetilde f\rangle =
   \widetilde\N \sum _{n=0}^\infty
   e^{-i \alpha \varepsilon_{2n}}  {\frac{\sqrt{[\varepsilon_{2n}]!}}{n!}}
   \left[\frac{\exp(i \phi)\tanh r}{2}\right]^n|2n\rangle,
 \end{equation}
  and
 \begin{equation}\label{DSS33}
   |\widetilde{\xi, \alpha, \widetilde f}\rangle =
    \widetilde\N' \sum _{n=0}^\infty e^{- i \alpha \varepsilon_{2n}}
     \frac{(2n)!}{n!\;\sqrt{[\varepsilon_{2n}]!}}
      \left[\frac{\exp(i \phi)\tanh r}{2}\right]^n|2n\rangle,
 \end{equation}
   where $\widetilde{\N}$ and  $\widetilde\N'$ may be
   determined by the normalization condition as follows,
  \begin{equation}\label{Norm}
    \widetilde \N =
     \left[ \sum_{n=0}^\infty \frac {[\varepsilon_{2n}]!}{(n!)^2}
      \left(\frac{\tanh r}{2}\right)^{2n} \right]^{-\frac{1}{2}},
  \end{equation}
 and
 \begin{equation}\label{Normp}
     \widetilde \N'=\left[ \sum_{n=0}^\infty \left( \frac {(2n)!}{n!} \right)^2
      \frac {1} {[\varepsilon_{2n}]!}  \left( \frac {\tanh r}{2}\right)^{2n}
       \right]^{-\frac 1 2}.
 \end{equation}
  respectively.
  The introduced states  in (\ref{SS11}) and  (\ref{SS22})
  (in (\ref{DSS22}) and (\ref{DSS33}))
  explicitly show the relation of
  Gazeau$-$Klauder squeezed states (the dual of Gazeau$-$Klauder squeezed states)
  to the spectrum of the quantum system (the dual of the quantum system).
  Note that from the form of the four classes of
  obtained squeezed states in equations  (\ref{SS11}), (\ref{SS22}),
 (\ref{DSS22}) and (\ref{DSS33}), it may be recognized that they are
  temporally  stable, i.e. possess one of the main features of the
  Gazeau$-$Klauder  coherent states. This is in fact due to the existence of the
  exponential term $\exp(\pm i\alpha \epsilon_n)$
  in the expansion coefficient of the obtained  squeezed states
  ($\epsilon$ stands appropriately for $e$ or $\varepsilon$).
  Hence, for instance by the following definition,
  the invariance of the  squeezed states under appropriate time evolution
  operator can be guaranteed, i.e.
 \begin{equation}\label{temp1}
   e^{i \HD t }|{\xi, \alpha, f}\rangle =
    |{\xi, \alpha +i\omega t, f}\rangle, \qquad
      e^{i \HD t }|\widetilde{\xi, \alpha, f}\rangle =
      |\widetilde{\xi, \alpha +i\omega t, f}\rangle,
\end{equation}
and
 \begin{equation}\label{temp3}
   e^{i\widetilde {\HD}t} |{\xi, \alpha, \widetilde f}\rangle=
    |{\xi, \alpha +i\omega t, \widetilde f}\rangle,\qquad
    e^{i\widetilde {\HD}t} |\widetilde{\xi, \alpha, \widetilde f}\rangle=
      |\widetilde{\xi, \alpha +i\omega t, \widetilde f}\rangle.
\end{equation}
   To investigate the above equations, there should be emphasis on using
   the eigen-value equations
   $\HD |n\rangle=e_n |n\rangle$ in (\ref{temp1}) and
   $\widetilde{\HD} |n\rangle=\varepsilon_n |n\rangle$ in (\ref{temp2}).
   Therefore, seemingly the name {\it "temporally stable squeezed states"} for the
   states  introduced in (\ref{SS11}), (\ref{SS22}), (\ref{DSS22}) and (\ref{DSS33})
   is suitable, if one chooses the
   normally ordered Hamiltonian composed from multiplication of
   annihilation and creation operators, respectively, in the evolution operator.

    Based on the nonlinear coherent states formalism in \cite{Manko1997},
    recently the {\it "nonlinear vacuum  squeezed states"} have been introduced
    through the following actions on the vacuum states \cite{Kwek}:
  \begin{equation}\label{NLSSs1}
     S(\xi)|0\rangle =\exp\left[\frac 1 2 (\xi {A^\dag}^2-\xi^*
     B^2)\right]|0\rangle=|\xi, f\rangle,
  \end{equation}
  \begin{equation}\label{NLSSs2}
    \tilde S(\xi)|0\rangle = \exp \left[\frac 1 2 (\xi
    {B^\dag}^2-\xi^* A^2)\right] |0\rangle=|\widetilde{\xi,
    f}\rangle\; .
\end{equation}
    In the above equations $A$ and $A^\dag$ may be obtained by
    simply setting $\alpha=0$ in (\ref{crea-anni}), similarly for $B$ and $B^\dag$.
    The number states expansion of the first one, $|\xi, f\rangle$ in (\ref{NLSSs1}), has
    the following superposition \cite{Kwek},
 \begin{equation}\label{SS-Kwek}
    |\xi, f\rangle =\N \sum _{n=0}^\infty \frac{\sqrt{(2n)!}}{n!}[f(2n)]!
    \left[\frac{\exp(i \phi)\tanh r}{2}\right]^n|2n\rangle,
 \end{equation}
   where $\xi= \tanh r \; \exp(i \phi)$ and $\N$ is chosen so that
   the states be normalized.
   Then the authors have studied the
   statistical properties of the  squeezed states (\ref{SS-Kwek}) for a special
   case with nonlinearity function describing the center of mass motion of a trapped
   ion (TI):
 \begin{equation}\label{f-TI}
     f_{\rm TI}(n)=L_n^1(\eta ^2)[(n+1)L_n^0(\eta^2)]^{-1},
 \end{equation}
   where $\eta$ is the Lamb-Dicke parameter and $L_m^n(x)$ are
   associated Laguerre polynomials.
   It can be easily investigated that the presented formalism recovers the results of
   \cite{Kwek} as a special case.
   Taking $f$ to be a real function, i.e.  setting $\alpha =0$
   in (\ref{SS1}) (or $e_n = nf^2(n)$ in (\ref{SS11})), eventually
   arrives one at the nonlinear  squeezed states in (\ref{SS-Kwek}).
   Also, it is notable that  setting $\alpha =0$
   in (\ref{DSS-Kwek}) (or $e_n = nf^2(n)$ in (\ref{SS22})),
   yields  the {\it "dual family of nonlinear squeezed states"}
   in (\ref{SS-Kwek}) as
 \begin{equation}\label{SS-Kwek-D}
    |\widetilde{\xi, f}\rangle = \N' \sum _{n=0}^\infty \frac{\sqrt{(2n)!}}{n!}
     \frac{1}{[f(2n)]!}
      \left[\frac{\exp(i \phi)\tanh r}{2}\right]^n|2n\rangle.
  \end{equation}
    The latter states are the number states expansion of (\ref{NLSSs2}).
    The states obtained in (\ref{SS-Kwek}) and (\ref{SS-Kwek-D}) are exactly
    equations (11a) and (11b) of a recent paper, respectively \cite{Obada}.

   Moreover, it ought to be mentioned here that the constructed squeezed states
   in this paper   have been called the "Gazeau$-$Klauder squeezed states" do not
   fully guarantee the criteria of Gazeau and Klauder \cite{gazklau}.
   The fact that might be expected.
   Since relaxing from the "action identity" criteria
   (which imposed on the Gazeau$-$Klauder coherent states
   in order to emphasize on the "classicality" of states) is neither necessary
   nor suitable
   here, because the squeezed states  are not essentially expected
   to show classical exhibition. Rather, generally most interesting in constructing
   the squeezed states is the nonclassical nature  of them.

\section{Gazeau$-$Klauder squeezed states of some physical systems}\label{examples}
  As some physical appearance of the proposed formalism, it is now possible to
  apply the scheme to a few well-known systems, i.e. simple harmonic
  oscillator, P\"{o}schl-Teller and the infinite  square-well potentials,
  hydrogen-like spectrum and at last the center of mass motion of a trapped ion.
  Gazeau$-$Klauder  coherent states and the corresponding dual pairs
  of all these systems (except the last one)
  have been previously constructed \cite{Roknizadeh-Tav-AIP}.

   {\it Example 1:  Harmonic oscillator}

   As the simplest example one can apply the formalism to the harmonic
   oscillator Hamiltonian, whose nonlinearity function is
   equal to $1$, hence $ \varepsilon_n = n=e_n$ and so $\mu(n)=n!=\rho(n)$.
   Note that we have considered a shifted Hamiltonian to lower the
   ground states energy to zero ($e_0=0=\varepsilon_0$). Eventually,
   it  can be easily observed that for the case of harmonic oscillator,
   all the four classes of the Gazeau$-$Klauder squeezed states coincide with each other
   in the following way:
 \begin{eqnarray}\label{SS-HO}
   |\xi, \alpha, f\rangle &=& \N \sum _{n=0}^\infty \frac{\sqrt{(2n)!}}{n!}
          \left[\frac{\exp(i \phi)\tanh
   r}{2}\right]^n
   e^{-i \alpha {2n}}|2n\rangle \nonumber \\
    &=& |\xi, \alpha, f\rangle = |\widetilde{\xi, \alpha, f}\rangle=
   |\xi, \alpha, \widetilde f\rangle =|\widetilde{\xi, \alpha, \widetilde f}\rangle.
\end{eqnarray}
   For these states the normalization constant can be
   evaluated in a closed form as follows: $ \N=(\cosh r)^{-\frac 1 2 }$.
   Relation (\ref{SS-HO}) clearly illustrates the
   {\it "self-duality"} of the Gazeau$-$Klauder squeezed states of harmonic
   oscillator.
   Ordinarily the self-duality,  holds in this case,
   can be viewed as a checkpoint to be sure about the presented formalism
   \cite{Roknizadeh2004, Roknizadeh-Tav-AIP}.
   Note that substituting $\alpha=0$ in (\ref{SS-HO}) will recover the exact form of the
   squeezed vacuum obtained by the unitary operator $S(\xi)$ in (\ref{ss}).
   Strictly speaking,  comparing $| \xi\rangle$  in (\ref{ss})
   and $| \xi, \alpha\rangle$ in (\ref{SS-HO}), it can be easily observed that $\xi$
   maps to $\xi \exp(-2 i \alpha)$, both in the complex
   plane, by the Gazeau and Klauder approach.


  {\it Example 2:  P\"{o}schl-Teller and infinite square-well potentials}
    These potentials and their coherent states are interesting due to
    various applications in
    many fields of physics such as atomic and molecular physics.
    The  Gazeau$-$Klauder coherent states corresponding to the
    P\"{o}schl-Teller potential, have been demonstrated by J-P. Antoine
    {{\it et al}} in \cite{Antoine2001}. Their obtained results are based
    on the eigenvalues
 \begin{equation}\label{poshCnd}
    e_n=n(n+\nu),  \qquad \nu > 2.
 \end{equation}
   In (\ref{poshCnd}) $\nu=\lambda+\kappa$, where $\lambda$ and $\kappa$
   are two parameters that determine the form
   (i.e. depth and width) of the potential well.
   Consequently, using the presented formalism the explicit form of the four
   classes of Gazeau$-$Klauder squeezed states corresponding to the
    P\"{o}schl-Teller potential read  as
 \begin{equation}\label{S-PT1}
   |\xi, \alpha, f\rangle = \N \sum _{n=0}^\infty
   e^{-i \alpha {2n(2n +\nu)}} \frac{(2n)!}{n!} \frac{\sqrt {[2n(2n
   +\nu)]!}}{n!} \left[\frac{\exp(i \phi)\tanh r}{2}\right]^n |2n\rangle,
 \end{equation}

 \begin{equation}\label{S-PT2}
   |\widetilde{\xi, \alpha, f}\rangle = \N'
   \sum _{n=0}^\infty
   e^{-i \alpha {2n(2n+\nu)}} \frac{(2n)!}{n!} \frac{(2n)!}{n!\sqrt {[2n(2n
   +\nu)]!}} \left[\frac{\exp(i \phi)\tanh r}{2}\right]^n|2n\rangle,
 \end{equation}

 \begin{equation}\label{S-PT3}
   |\xi, \alpha, \widetilde f\rangle =
   \widetilde \N \sum _{n=0}^\infty
   e^{-i \alpha \frac{2n}{ 2n+\nu}} \frac{1} {n!}
   \sqrt{ \left[\frac{2n}{2n+\nu}\right]!}
    \left[\frac{\exp(i \phi)\tanh r}{2}\right]^n|2n\rangle,
 \end{equation}

\begin{equation}\label{S-PT4}
  |\widetilde{\xi, \alpha, \widetilde f}\rangle =
  \widetilde \N' \sum _{n=0}^\infty
  e^{-i \alpha {\frac {2n}{2n+\nu}}}
  \frac{(2n)!}{n!}
  \sqrt { \left[\frac{2n +\nu}{2n}\right]!}
    \left[\frac{\exp(i \phi)\tanh r}{2}\right]^n|2n\rangle.
\end{equation}
  In the same manner, the squeezed states associated with the infinite
  square-well potential
  may be obtained  by replacing $\nu = 2$ in equations
  (\ref{S-PT1})-(\ref{S-PT4}).

{\it Example $3$   Hydrogen-like spectrum}
     We now choose the hydrogen-like
    spectrum whose the corresponding coherent states, has been a long-standing
    subject and discussed frequently in the literature.
    The eigen-values of the one-dimensional
    model of such a system with the Hamiltonian $\hat H = -\omega/(\hat n +1)^2$
    has been considered in \cite{gazklau} with eigenvalues
 \begin{equation}\label{hyd}
    e_n = 1- \frac{1}{(n+1)^2},
 \end{equation}
   to be such that $e_0=0$ (while $\omega=1$). So,
   the Gazeau$-$Klauder squeezed states   for this system  can be easily calculated as
 \begin{eqnarray}\label{SH-1}
   |\xi, \alpha, f\rangle &=& \N \sum _{n=0}^\infty
   e^{-i \alpha \frac{2n(2n+2)}{(2n+1)^2}}
   \frac{1}{n!}
   \sqrt {\left[ \frac{2n(2n+2)}{(2n+1)^2} \right]!} \nonumber \\
   &\times&
   \left[\frac{\exp(i \phi)\tanh r}{2}\right]^n|2n\rangle,
 \end{eqnarray}

 \begin{eqnarray}\label{SH-2}
   |\widetilde{\xi, \alpha, f}\rangle &=& \N' \sum _{n=0}^\infty
   e^{-i \alpha \frac{2n(2n+2)}{(2n+1)^2}}\frac{(2n)!}{n!}
   \sqrt {\left[\frac {(2n+1)^2}{2n(2n+2)}\right]!}
   \nonumber \\
   &\times&
   \left[\frac{\exp(i \phi)\tanh r}{2}\right]^n|2n\rangle,
 \end{eqnarray}

 \begin{eqnarray}\label{SH-3}
   |\xi, \alpha, \widetilde f\rangle &=&
   \widetilde \N \sum _{n=0}^\infty
   e^{-i \alpha \frac{2n(2n+1)^2}{2n+2}} \frac{1}{n!}
   \sqrt {\left[\frac{2n(2n+1)^2}{2n+2}\right]!}
   \nonumber \\
   &\times&
   \left[\frac{\exp(i \phi)\tanh r}{2}\right]^n|2n\rangle,
 \end{eqnarray}

 \begin{eqnarray}\label{SH-4}
   |\widetilde{\xi, \alpha, \widetilde f}\rangle &=&
   \widetilde \N' \sum _{n=0}^\infty
   e^{- i \alpha  \frac{2n(2n+1)^2}{2n+2}}  \frac{(2n)!}{n!}
   \sqrt {\left[\frac{2n+2}{2n(2n+1)^2}\right]!}
   \nonumber \\
   &\times&
   \left[\frac{\exp(i \phi)\tanh r}{2}\right]^n|2n\rangle.
 \end{eqnarray}
   Seemingly,
   this is the first time that the  squeezed states
   associated with the hydrogen-like atom are introduced
   in such a direct relation to
   the general structure of squeezed states and also to the related spectrum.

   {\it Example 4:  Center of mass motion of a trapped ion:}

   As a final example, the center of mass motion of a trapped ion
   with the nonlinearity function in (\ref{f-TI}) will be considered here.
   The associated (nonlinear) coherent and  squeezed states
   were of much interest in recent decade \cite{Matos1996, MMatos1996}.
   Fortunately, the presented formalism in section 2 allows one to define
   a $\hat n-$ dependent Hamiltonian associated with the trapped ion system
   such that
 \begin{equation}\label{HTI}
    \hat H_{\rm TI}= \hat n f_{\rm TI}^2(\hat n)= \frac{\hat n}{(\hat
    n +1)^2}\left[\frac {L_{\hat n}^1(\eta ^2)} {L_{\hat
    n}^0(\eta^2)}\right]^2,
 \end{equation}
  where we have used the nonlinearity function of trapped ion
  introduced in (\ref {f-TI}).
   It seems that the Gazeau$-$Klauder type of squeezed states
   is also possible to be introduced, if on considers
   the system with the $n$-dependent Hamiltonian as stated in (\ref{HTI}).
   Therefore, the system will be specified with the spectrum
  \begin{equation}\label{HTI-n}
     e_n = \frac{ n}{(n +1)^2}
       \left[\frac {L_n^1(\eta ^2)} {L_n^0(\eta^2)}\right]^2,
  \end{equation}
   from which the spectrum of the dual system will be easily calculated using
   $\varepsilon_n=\frac{n^2}{e_n}$. Hence, having $e_{2n}$ and $\varepsilon_{2n}$,
   the four classes of Gazeau$-$Klauder squeezed states
   for the center of mass motion of trapped ion may be
   easily obtained with the help of the general structure introduced in equations
   (\ref{SS11}), (\ref{SS22}), (\ref{DSS22}) and (\ref{DSS33}).

\section{The quantum statistical properties and squeezing of Gazeau$-$Klauder squeezed states}
    The quantum statistical properties of the squeezed states outlined in the present paper as well
    as the squeezing exhibition of them will be considered in this section. As one of the manifest
    nonclassicality features of all the generalized squeezed states obtained in this paper, one may
    refer to the 'oscillatory number distribution' of these states. Photon statistics of the states in
    (\ref{SS11}), (\ref{SS22}), (\ref{DSS22}) and (\ref{DSS33})
    may be easily obtained as
 \begin{equation}\label{}
    P(2n)=\N^2\frac{[e_{2n}]!}{(n!)^2}\left[\frac{\tanh r}{2}\right]^{2n},
 \end{equation}

 \begin{equation}\label{}
    P'(2n)=\N'^2 \left[\frac{(2n)!}{n!}\right]^2
    \frac{1}{[e_{2n}]!}\left[\frac{\tanh r}{2}\right]^{2n},
 \end{equation}

 \begin{equation}\label{}
   \widetilde P(2n)=
    \widetilde\N^2\frac{[\varepsilon_{2n}]!}{(n!)^2}\left[\frac{\tanh r}{2}\right]^{2n},
 \end{equation}

 \begin{equation}\label{}
   \widetilde P'(2n)=  \widetilde \N'^2 \left[\frac{(2n)!}{n!}\right]^2
    \frac{1}{[\varepsilon_{2n}]!}\left[\frac{\tanh r}{2}\right]^{2n},
 \end{equation}
  respectively.
  $\N$, $\N'$, $\widetilde\N$ and
    $\widetilde\N'$ in the above four relations  determined in
    (\ref{norm}), (\ref{normp}), (\ref{Norm})
    and (\ref{Normp}), respectively.
  Generally, it is seen that for all of the introduced Gazeau$-$Klauder  squeezed states one has
    \begin{equation}\label{dist}
     \mathbf P(2n)\neq 0, \quad{\rm while}
      \quad \mathbf P(2n+1)=0, \quad{\rm for\; all}\; n\;.
  \end{equation}
  which clearly shows the non-classicality
  nature of the obtained squeezed states in (\ref{SS11}), (\ref{SS22}), (\ref{DSS22}), (\ref{DSS33}).

   To complete  the study of the statistical properties of the squeezed states associated
   with the physical examples introduced in the previous section,
   the Mandel parameter and the
   squeezing of the quadratures of the field will be illustrated
   numerically, since the analytical form of the above quantities can
   not be given in closed form. The calculation of the Mandel
   parameter defined as:
\begin{equation}\label{Mandel}
  Q=\frac{\langle n^2\rangle - \langle n\rangle ^2}{\langle
  n\rangle}-1, \qquad n=a^\dag a,
\end{equation}
   determines the supper-Poissinian (if $Q > 0$), sub-Poissinian (if
   $Q < 0$) and Poissinian (if $Q = 0$) nature of the states. The case of the Poissinian
   is the characteristics of the standard coherent states. The sub-Poissinian
   is an important property which implies the non-classicality of the
   states, and the supper-Poissinian statistics has important
   consequence for the properties of localization and temporal
   stability of the wave packet \cite{Antoine2001}. For the squeezing of
   the states according to $x=(a+a^\dag)/\sqrt 2$, $p= =(a-a^\dag)/ (i \sqrt
   2$) one has to calculate the following quantities:
\begin{equation}\label{delp}
  (\bigtriangleup p)^2 = \langle p^2\rangle - \langle p\rangle ^2
                       = \langle a^\dag a\rangle - \frac 1 2 \langle a^2\rangle
                       - \frac 1 2 \langle (a^\dag) ^2\rangle  + \frac 1 2,
\end{equation}
\begin{equation}\label{delx}
  (\bigtriangleup x)^2 = \langle x^2\rangle - \langle x\rangle ^2
                       = \langle a^\dag a\rangle + \frac 1 2 \langle a^2\rangle
                       + \frac 1 2 \langle (a^\dag) ^2\rangle  + \frac 1
                       2.
\end{equation}
    In the latter two equations it is set $\langle a \rangle = 0 =
    \langle (a^\dag)\rangle$
   which holds for all classes of the obtained
    squeezed states (since all of them are some superpositions of the even Fock states, $| 2 n\rangle$).
   Note that all of the expectation values in  the equations (\ref{Mandel}), (\ref{delp}),
   (\ref{delx})
   must be calculated with respect to the squeezed states in states  in (\ref{SS11}) and  (\ref{SS22})
   (in (\ref{DSS22}) and (\ref{DSS33})) for the quantum physical
   examples of outlined in this paper. Squeezing holds in the $x$-, $p$-quadrature,
   if $(\bigtriangleup x)^2$,  $(\bigtriangleup p)^2$ be less
   than $\frac 1 2$, respectively.
   Seemingly, it will be enough to present here
   the needed terms for one of the squeezed states, other cases may be derived
   in a similar fashion.
   For the the states in (\ref{SS11}) it can be easily seen that,
  \begin{equation}\label{}
    \langle (a^\dag a) \rangle = 2 \N ^2 \sum _{n=0}^\infty \frac {[e_{2n+2}]!}{n!
    (n+1)!}\left ( \frac {\tanh r}{2}\right)^{2n+2},
  \end{equation}
  \begin{equation}\label{}
    \langle (a^\dag a)^2 \rangle = 4 \N ^2 \sum _{n=0}^\infty \frac {[e_{2n+2}]!}{
    [(n+1)!]^2}\left ( \frac {\tanh r}{2}\right)^{2n+2},
  \end{equation}

\begin{eqnarray}\label{}
    \langle a^2 \rangle &=&  \N ^2 \sum _{n=0}^\infty
     e^{i \alpha (e_{2n}-e_{2n+2})}\\ \nonumber
     &\times&
     \frac {\sqrt {(2n+1)(2n+2) [e_{2n+2}]! [e_{2n}]!}} {n!
    (n+1)!} e^{-i \phi} \left ( \frac {\tanh r}{2}\right)^{2n+1},
  \end{eqnarray}

\begin{eqnarray}\label{}
    \langle {a^\dag}^2 \rangle = \langle {a}^2 \rangle^ \dag.
  \end{eqnarray}

   Setting $\phi=0$ and the  eigenvalue $e_{2n}$ associated
   with the physical examples of section \ref{examples}, $Q, (\triangle x)^2,
   (\triangle p)^2$ can be easily evaluated. Similar calculations can be done for the states
   in (22), (28) and (29), straightforwardly.

Part of the numerical calculations, for some classes of the
Gazeau$-$Klauder squeezed states have been presented in figures which
follows. Figure 1a shows the super-Poissonian statistics for S.H.O.
(simple harmonic oscillator) for all values of r (note that the
results for any state of S.H.O. cover all the four classes of
squeezed states for S.H.O, due to its self-duality).
For the trapped ion system the states in (\ref{SS11}) shows the
sub-Poissonian statistics for a wide rang of values of r and $\eta$
(figure 1b shows $Q$ as a function of $r$ for fixed value of
$\eta=0.5$ and figure 1c shows $Q$ versus $\eta$ for
fixed value of $r=1$). Figure 1d describes $Q$ versus $r$ with the
choice $\eta = 0.7$ for the trapped ion corresponding to formula
(\ref{DSS33}). It illustrates the supper-Poissonian exhibition of
the constructed squeezed states in a wide range of the values of
$r$.

The numerical calculations show that the Gazeau$-$Klauder type of squeezed states of
trapped ion motion according to equation (22) is super-Poissonian, using the parameters
$r \geq 0.1, \eta = 0.7$, and the same system according to equation (28) has $Q < 0$, for
$r \geq 0.03, \eta = 0.7$. The Mandel parameter for infinite square-well and P\"{o}schl-Teller potentials
gives $Q < 0$ when equation (19) has been used, $Q > 0$ for equation (22), $Q > 0$ while
equation (28) has been in consideration and $Q < 0$ for equation (29). It is worth noting
that generally all of the sub-Poissonian cases whose results have been given here without
the graphs have $Q > 0$ very near to $r\simeq 0$.  The calculations
of $Q$ for the Hydrogen atom shows $Q>0$ for equation (\ref{SS11}),
$Q<0$ for equation (\ref{SS22}) $Q<0$ for equation (\ref{DSS22}) and
$Q>0$ when equation (\ref{DSS33}) has been used.

The plot of squeezing in $x$- and $p$-quadrature for the S.H.O. has
been shown in Figure 2a in terms of the fixed value of $r=1$.
Squeezing can be observed for $x$-quadrature,  when $1.22 \leq
\alpha \leq 1.92$, and also in  $p$-quadrature in two distinct
regions: when and $0 \leq \alpha \leq 0.35$ and $2.8 \leq \alpha
\leq 3.49$.
Again, for
the harmonic oscillator $(\bigtriangleup x)^2$ is plotted as a function of $r$ for a fixed value of $\alpha = 1.5$ in
figure 2b. It is observed that the squeezing occurs for $r \leq 2.6$ in $x$-quadrature.
Figure 2c indicates the squeezing in $p$-quadrature for the hydrogen atom according to formula (19)
$\bigtriangleup x$ as a function of
$r$  (when $\alpha=1.5$) which occurs for all values of $r$. It is shown that the squeezing
occurs for $r\leq 2.6$ in $x$-quadrature. Figure 2c indicates the
squeezing in $p$-quadrature for the Hydrogen atom according to the
state (\ref{SS11}) (when   $\alpha=1.5$)  which occurs for all values
of $r$.

Figure 3a demonstrates the squeezing in $p$-quadrature of the
P\"{o}schl-Teller ($\nu=5$) for $\alpha=4$  and potential well
($\nu=2$) for  $\alpha=0.1$, when the equation (\ref{SS22}) has been used.
In this case the squeezing may interpolate between $x$- and $p$-quadrature by
tuning $\alpha$ and $r$ parameters. In figure 3b $\triangle x$ of
the trapped ion of equation (\ref{SS22}) has been plotted,  the squeezing
of which may be observed in the range $r\leq 0.5$ (the other parameters are choosen
so that $ \eta= 0.1$ and $\alpha=1.5$).

In figure 4a the squeezing in $x$-quadrature for the trapped ion
of the equation (\ref{DSS22}) is plotted (using the parameters $
\eta= 0.3$ and $\alpha=1.5$). As can be observed, squeezing occurs
in the range $r\leq 0.6$. Figure 4b indicates the squeezing in $p$-quadrature 
of the P\"{o}schl-Teller ($\nu=20$)  and potential well
($\nu=2$), where in both cases  the $\alpha$ parameter is choosen to be
$1.5$.
It is interesting to note that the squeezing occurs in the whole range of $r$ for
the two cases, and the one for the infinite square-well potential is always
stronger. at last, the numerical calculations for the Hydrogen atom in
equation (\ref{DSS33}) shows that the $p$-quadrature is squeezed for all
values of $r$, when $\alpha$ is choosed equal to $1.5$ (figure
5).
Altogether, by the above results the non-classicality nature of the introduced
squeezed states in this paper has been established, obviously.

 {\bf Acknowledgments:}
 The author would like to express his utmost thanks to Dr R Roknizadeh at the Quantum Group
 of the University of Isfahan for his insightful comments. Also, he is pleased to acknowledge
 the referees for their useful suggestions which considerably improved the quality of the paper.
 Finally, thanks to the research council of the University of Yazd for financial supports to this
 project.

\vspace{2cm}

{\bf Figure Captions:}

Fig. 1a: {Mandel parameter of Gazeau$-$Klauder  squeezed states of harmonic oscillator as a
   function of $r$.} \label{}

Fig. 1b: {Mandel parameter of trapped ion of Gazeau$-$Klauder  SSs of equation
   (\ref{SS11}) as a function of $r$ ($\eta= 0.5$).} \label{}

Fig. 1c: {Mandel parameter of trapped ion of Gazeau$-$Klauder  SSs of equation
   (\ref{SS11}) as a function of $\eta$ ($r= 1$).} \label{}

Fig. 1d: {Mandel parameter of trapped ion of Gazeau$-$Klauder  SSs of equation
   (\ref{DSS33}) as a function of $r$ ($\eta= .7$).} \label{}

Fig. 2a: {The squeezing in $x$-quadrature(solid line)
 and $p$-quadrature(dashed line)
   for harmonic oscillator as a function of $\alpha$ ($r=1$).}

Fig. 2b:  {Plot of ($\triangle x$) as a function of $r$
   ($\alpha=1.5$) for harmonic oscillator.} \label{}

Fig. 2c:  {Plot of ($\triangle p$) as a function of $r$ ($\alpha=.5$)
   of the Hydrogen atom, when the states of equation (\ref{SS11}) has been
   used.} \label{}

Fig. 3a: {Plot of ($\triangle p$) as a function of $r$ for the
   potential well, $\nu=2$, (solid line) and P\"{o}schl-Teller
   potential, $\nu=5$, (dashed line). The structure of equation
   (\ref{SS22}) is considered.} \label{}
   
Fig. 3b: {Plot of $(\triangle x)$ of the trapped ion system of the Gazeau$-$Klauder  squeezed states
   of equation (\ref{SS22}) as a function of $r$
   ($\alpha=1.5, \eta=.1$).} \label{}

Fig. 3c: {Plot of $(\triangle x)$ of the trapped ion system of the Gazeau$-$Klauder  squeezed states
   of equation (\ref{DSS22}) as a function of $r$
   ($\alpha=1.5, \eta=.3$).} \label{}

Fig. 4a: {Plot of $(\triangle p)$ as a  function of $r$ for
    potential well ($\nu=2$) and P\"{o}schl-Teller potential
    ($\nu=20$),
    where the equation (\ref{DSS22}) is used and $\alpha=1.5$.} \label{}
    
Fig. 4b: {Plot of squeezing in $p$-quadrature as a function
    of $r$ for the Hydrogen atom where the equation (\ref{DSS33}) and $\alpha=1.5$
    is used.} \label{}


\include{thebibliography}

   \end{document}